\documentclass[aps,prb,twocolumn,showpacs,superscriptaddress,]{revtex4-1}
\usepackage[latin1]{inputenc}
\usepackage[english]{babel}
\usepackage{graphicx, color}
\usepackage{hyperref}
\definecolor{link}{rgb}{0.1,0.1,0.9}
\hypersetup{colorlinks=true,linkcolor=link,citecolor=link,urlcolor=link,linktocpage}
\usepackage{epstopdf}
\usepackage{color}
\usepackage{amsmath}
\usepackage{amssymb}
\usepackage{longtable}
\usepackage{mathtools}
\usepackage{float}

\begin{document}
	
\title{A magnetocaloric study of the magnetostructural transitions in NiCr$_2$O$_4$}

\author{Anzar Ali} 
\affiliation{Department of Physical Sciences, Indian Institute of Science Education and Research, Knowledge city, Sector 81, SAS Nagar, Manauli PO 140306, Mohali, Punjab, India}

\author{Gyaneshwar Sharma}
\affiliation{Department of Physical Sciences, Indian Institute of Science Education and Research, Knowledge city, Sector 81, SAS Nagar, Manauli PO 140306, Mohali, Punjab, India}
\affiliation{Department of Applied Sciences, PEC University of Technology, Sector 12, Chandigarh 160012, India}

\author{Yogesh Singh} \email{yogesh@iisermohali.ac.in}
\affiliation{Department of Physical Sciences, Indian Institute of Science Education and Research, Knowledge city, Sector 81, SAS Nagar, Manauli PO 140306, Mohali, Punjab, India}

\begin{abstract}
The spinel NiCr$_2$O$_4$ is known to show a ferrimagnetic transition at $T_c = 70$~K, and magneto-structural transitions at $T_s = 30$~K and $T_o = 20$~K\@.  We present a detailed magnetic and magnetocaloric effect (MCE $= -\Delta S_{M}(T)$) study across these transitions.  The $-\Delta S_{M}(T)$ shows a positive anomaly at $T_c$, $T_s$, and $T_o$.  In addition to these anomalies, we report a new unreported feature at $T \approx 8.5$~K where $-\Delta S_{M}(T)$ shows a negative anomaly or the inverse MCE.  An Arrot plot of the isothermal magnetization data reveals important information about the nature of the possible phases revealed in $-\Delta S_{M}(T)$.  We have also made a scaling analysis of the $-\Delta S_{M}(T)$ data around these transitions.  This analysis suggests that the transition at $T_c$ is a second-order Mean field like transition, the transition at $T_s$ is not second order and is non-mean field like, while the new transition at $T = 8.5$~K is non-mean field like but is second order in nature.  Our study demonstrates that magnetocaloric effect is sensitive to magneto-structural changes in materials and can be used for the identification of new phases and transitions.  

\end{abstract}

\maketitle

\section{Introduction}

Materials crystallizing in the spinel structure show many multifunctional behaviours such as, magnetodielectric, multiferroic, magnetoelastic, and magnetocaloric effects.  The variety of such multifunctional behaviours in a single material class is due to correlations among different degrees of freedom such as spin, lattice and orbital \cite{1,2,3,4,5,6,7,8,9,10,11,12}. The spinel oxides are any of a class of minerals with the chemical formula $A^{2+}B^{3+}_2$O$_4$ \cite{1,12,13,14,15,16}.   NiCr$_{2}$O$_{4}$ is a normal spinel material crystallizing in the cubic space group $Fd\bar{3}m$ at high temperatures \cite{3,18}.  As the temperature is lowered NiCr$_2$O$_4$ undergoes multiple transitions.  A structural transition from cubic to tetragonal symmetry occurs at $310$~K \cite{20,21}. A coupled magneto-structural change from tetragonal to orthorhombic symmetry accompanied by ferrimagnetic order between the Cr and Ni sublattices occurs at $T_c \approx 70$~K \cite{22,23}.  Another magneto-structural transition occurs at $T_s = 30$~K involving subtle changes in $NiO_{4}$ tetrahedra \cite{21,22,24,25,26}.  Recently another previously unreported feature at $T_o = 20$~K has been identified using magnetocapacitance measurements \cite{19}.  This has been suggested to be another coupled magneto-structural change where completion of ferrimagnetic ordering and a spin-driven structural distortion occurs.  Thus NiCr$_2$O$_4$ has been shown to be an avenue to study effects of strong coupling between the spin and lattice degrees of freedom.   

The magnetocaloric effect (MCE) has been recently proposed to be a sensitive measurement to study magnetoelastic systems where the magnetic and lattice degrees of freedom are strongly coupled \cite{2, 30}.  NiCr$_2$O$_4$ therefore appears to be an ideal material for a magnetocaloric study to explore whether the various magneto-structural transitions give measurable response in MCE.  Here we present a detailed study of the temperature and magnetic field dependence of the magnetization $M$ and the derived magnetocaloric response on polycrystalline samples of NiCr$_2$O$_4$.  The $\frac{dM(T,H)}{dT}$ curves show anomalies at $T = 70$~K and $30$~K suggesting these transitions have a magnetic component to them. An Arrott plot is used to get insight into the nature of these transitions.  The magneto caloric effect (MCE = $-\Delta S_M$) shows clear positive anomalies at $T_c = 70$~K and $T_s = 30$~K as well as at $T_o = 20$~K thus confirming that some magneto-structural change occurs at $20$~K\@.  In addition to these anomalies, we observe a prominent negative anomaly in $-\Delta S_M$ (inverse MCE) at $T = 8.5$~K pointing to an as yet undiscovered phase change in NiCr$_2$O$_4$.  Thus the magnetocaloric effect indeed seems to be a sensitive probe to detect magneto-structural changes in materials.

\section{Experiment}
The polycrystalline samples of NiCr$_2$O$_4$ were prepared by conventional solid state reaction as reported previously \cite{19}.  Stoichiometric amounts of NiO ($99.995~\%$ Alfa Aesar) and Cr$_{2}$O$_{3}~(99.999~\%$ Alfa Aesar) were mixed in an agate mortar and pestle for approximately $1$~hour. This mixture was pelletized and sintered at $800$~\textdegree C for $12$~hours in air.  The resulting pellets were reground, mixed, re-pelletized, and annealed at $1100$~\textdegree C for $24$~hours. This step was repeated twice. The X-ray powder diffraction data were collected at room temperature and analyzed using GSAS software \cite{27}. The magnetic measurements were performed using a physical property measurement system from Quantum Design in the temperature range  $2K \le T \le 300K$ and in magnetic fields in the range $0 \le H \le 9$~T\@.

\begin{figure}
	\centering
	\includegraphics[width=1.05\linewidth]{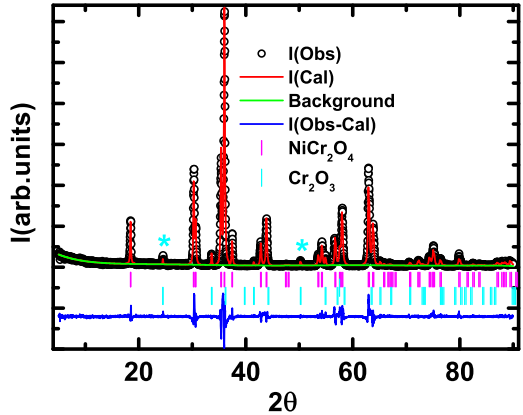}
	\caption{Rietveld refinement of room temperature X-ray diffraction pattern of NiCr$_2$O$_4$.  Bragg peak of the impurity phase Cr$_2$O$_3$ is labeled with $*$.  Upper (Magenta) bars ($|$) and lower (cyan) bars ($|$) represent the expected Bragg positions for NiCr$_2$O$_4$ and Cr$_2$O$_3$, respectively.  The difference curve betwen the experimental pattern and the refined pattern is given at the bottom.}
	\label{XRD}
\end{figure}

Figure~\ref{XRD} shows the powder X-ray diffraction (PXRD) pattern of the synthesized NiCr$_2$O$_4$ material.  The Bragg peaks in this pattern indicate that NiCr$_2$O$_4$ crystallizes in the expected tetragonal space-group $I4_{1}/amd$ consistent with previous literature \cite{22}.  X-ray diffraction data shows presence of a small amount ($\leq 5\%$) of Cr$_2$O$_3$ impurity phase.  A Rietveld refinement of the PXRD data was done.  The results of the fit, shown in Fig.~\ref{XRD}, gave lattice parameters close to those previously reported for NiCr$_2$O$_4$ \cite{22}.

\section{Results and Discussion}
\begin{figure}
	\includegraphics[width=.85\linewidth]{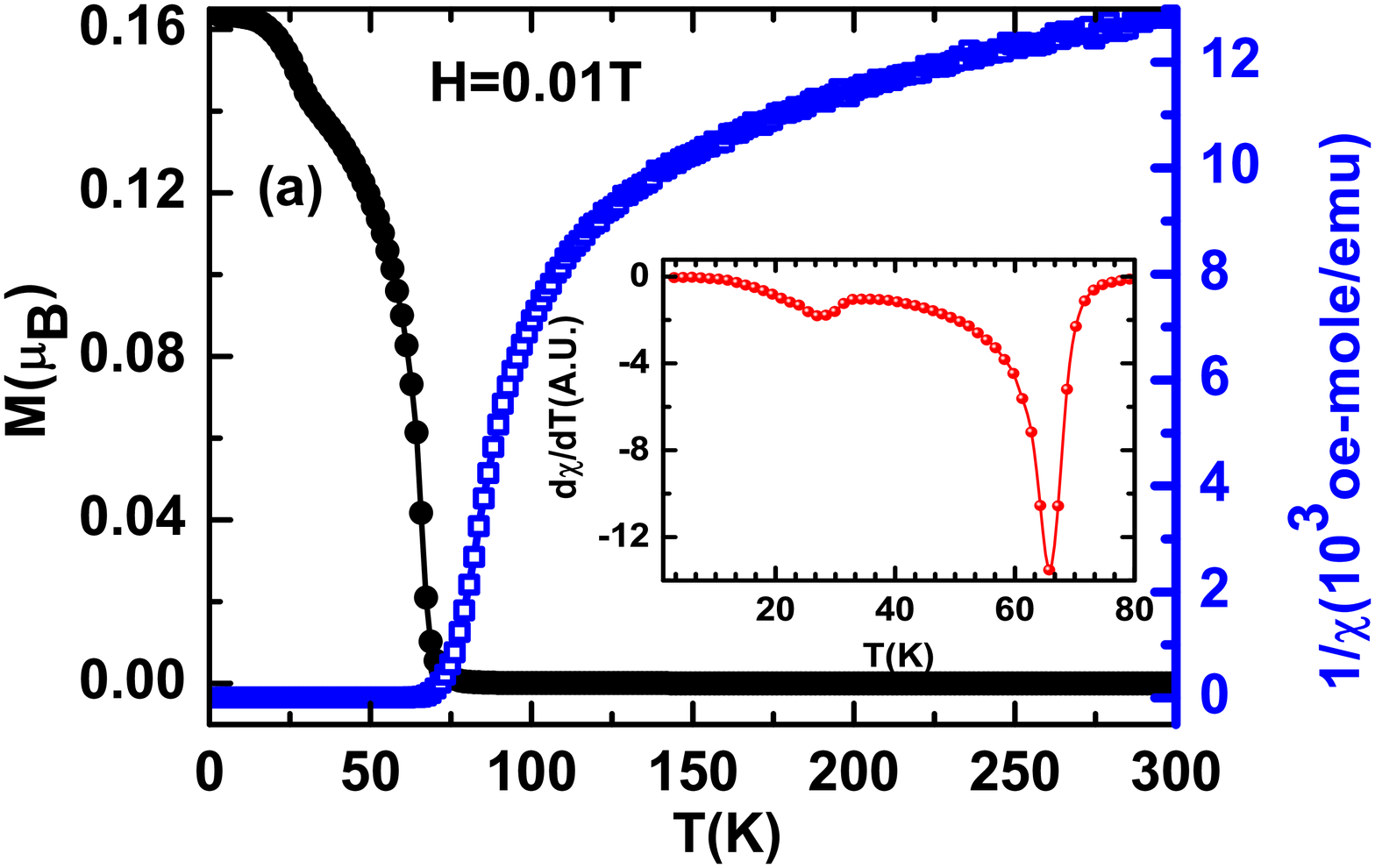}
	\includegraphics[width=.85\linewidth]{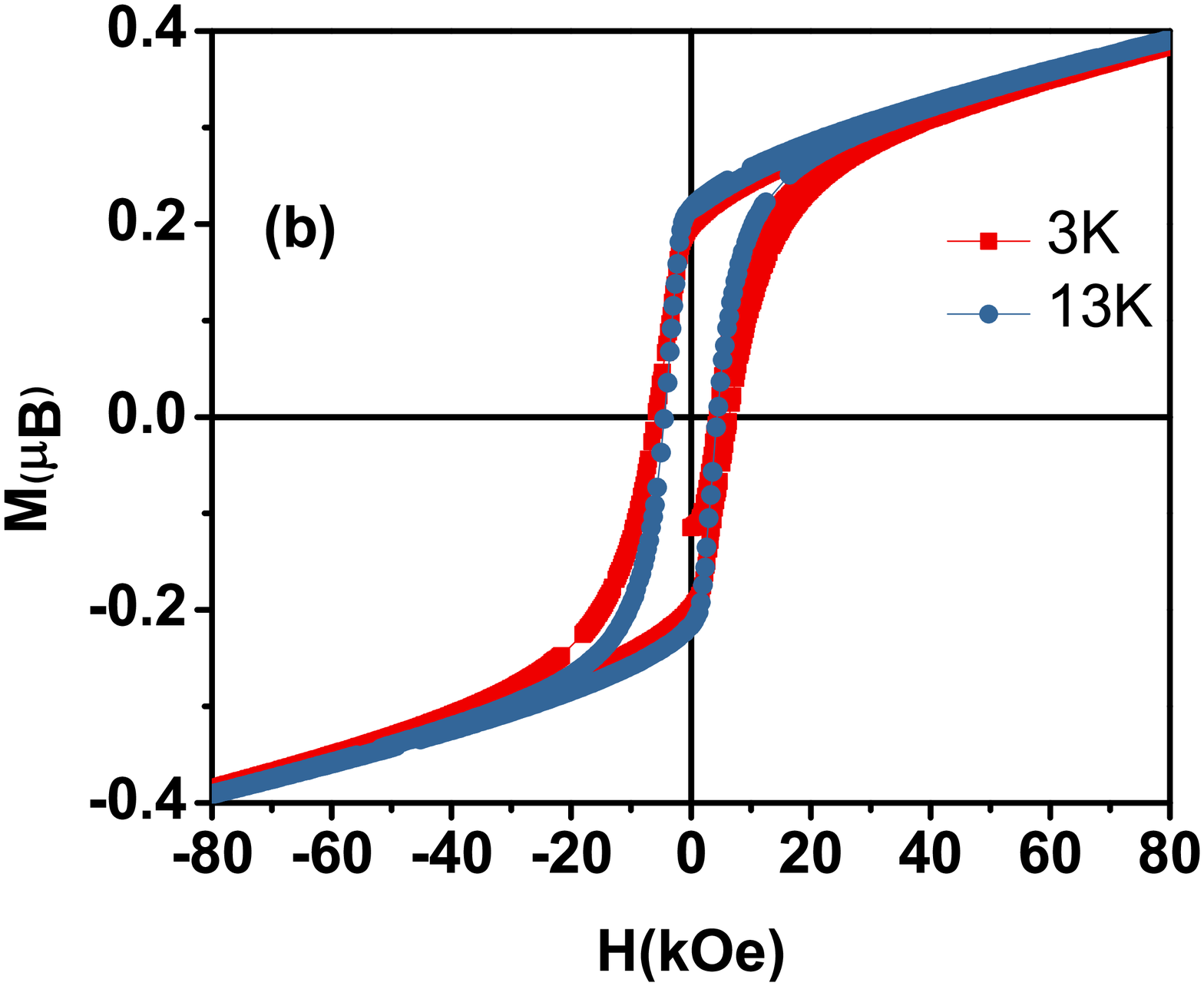}
	\caption{Magnetization vs temperature of NiCr$_2$O$_4$ in a magnetic field of $0.01$~T on the left $y$-axis. The inverse magnetic susceptibility $1/\chi$ on the right $y$- axis approaches $T_c$ from he paramagnetic phase with a hyperbolic $T$ dependence. Inset, temperature derivative of $\chi$ vs temperature shows two clear anomalies.}
	\label{CW}
\end{figure}

\subsection{Temperature and field dependent magnetization}

   \begin{figure}
   	\includegraphics[width=.85\linewidth]{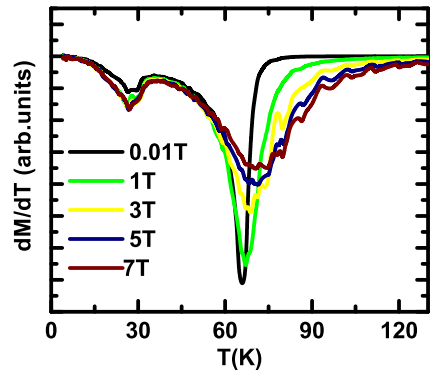}
   	\caption{The $dM/dT$ versus $T$ at various magnetic fields.}
   	\label{fig:mth-dispersion}
   \end{figure}

Figure \ref{CW} shows $M$ vs $T$ (on the left $y$-panel) and inverse susceptibility, $1/\chi(T) = H/M$ (on right $y$-panel) measured in a magnetic field $H = 0.01$~T\@.  The $M(T)$ shows a sharp increase to very large values on cooling below $T_c = 70$~K indicating a transition to a ferromagnetic or ferrimagnetic phase. Below $T = 30$~K the magnetization shows another anomaly where the magnetization decreases suggesting the onset of a further transition, possibly antiferomagnetic in nature.  These two anomalies are more clearly visible in the $\frac{d\chi}{dT}$ plot shown in the inset of Fig.~\ref{CW}.  The $1/\chi(T)$ follows a linear $T$ dependence at high temperatures but shows a deviation from linear behavior as $T_c$ is approached from above.  Near $T_c$, this nonlinearity in $1/\chi(T)$ becomes hyperbolic in nature.  A fit above $T = 200$~K to the expresion $\chi(T) = \chi_0 + C/(T - \theta)$, where $C$ is the Curie constant and $\theta$ is the Weiss temperature, gave the values $C = 6.33$~cm$^3$/mole~K and $\theta = -604$~K\@.  The large and negative $\theta$ implies predominantly antiferromagnetic exchange among magnetic ions.  NiCr$_2$O$_4$ is thus a geometrically frustrated magnet with a frustration index $f = \theta/T_N \approx 10$.

Figure \ref{CW} shows the field dependent isothermal magnetization data at temperatures $T = 3$ and $13$~K inside the magnetically ordered state.  A rapid increase in $M$ upto $H = 2$~T followed by a much weaker, almost linear increase for larger $H$ is observed.  Additionally, there is a hysteresis between the $M(H)$ data measured in increasing and decreasing $H$.  The value of $M$ at the highest fields is much smaller than expected for completely polarized NiCr$_2$O$_4$.  The above results  i.e. $\theta <0$, $\frac{dM}{dT}<0$, hyperbolic behavior of $1/\chi(T)$ \cite{23} when $T \rightarrow T_c$, and the $M(H)$ behaviour is typical of a  paramagnetic to ferrimagnetic transition \cite{28}.
The above results including the two phase transitions are consistent with previous reports on this material \cite{19, 23, 26, 29}. 
       
\begin{figure}
   	\includegraphics[width=0.85\linewidth]{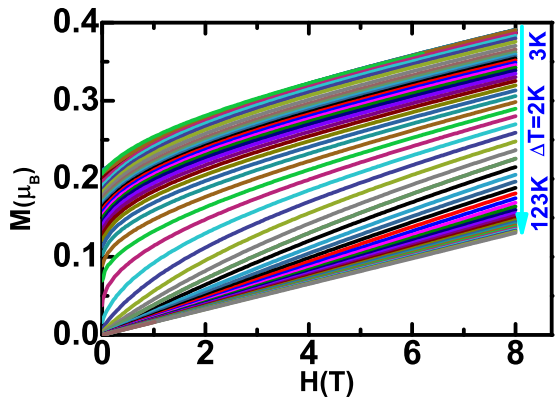}
   	\caption{Isothermal magnetization versus $H$ measured from $T = 3$~K to $123$~K every $2$~K.}
   	\label{Series_M(H)}
   \end{figure}

To get further insight into the various observed anomalies, we look in more detail at the temperature derivative of magnetization $dM/dT$ at various magnetic fields as shown in Fig. \ref{fig:mth-dispersion}. The anomaly in $dM/dT$ near $T_c$ shows a stronger dependence on magnetic field.  The position of the anomaly moves up in temperature and its gets broadened in larger fields.  This strong field dependence strongly indicates a magnetic origin for the transition at $T_c$.  On the other hand, the position of the anomaly at $T_s = 30$~K is almost independent of magnetic field and it becomes slightly sharper in larger fields.  This weak field dependence suggests a mostly structural origin for the transition at $T_s$.  

  \subsection{Magnetocaloric effect} 
The MCE is given by the negative of the magnetic entropy change $-\Delta S_M(T,H)$ and can be derived from the isothermal magnetization at various $T$ and $H$ using the expression \cite{2},
    
   \begin{equation}
   \Delta S_M (T, H_{0 \rightarrow H_{MAX}}) = \int _{0}^{H_{MAX}} \left| \dfrac{dM}{dT}\right|_H dH.
   \label{Maxwell}
   \end{equation}
If magnetization data is available only at discrete values of temperature, as is mostly the case in experiments, the integral in the above expression is replaced by a summation.

\begin{figure}
   	\includegraphics[width=1.05\linewidth]{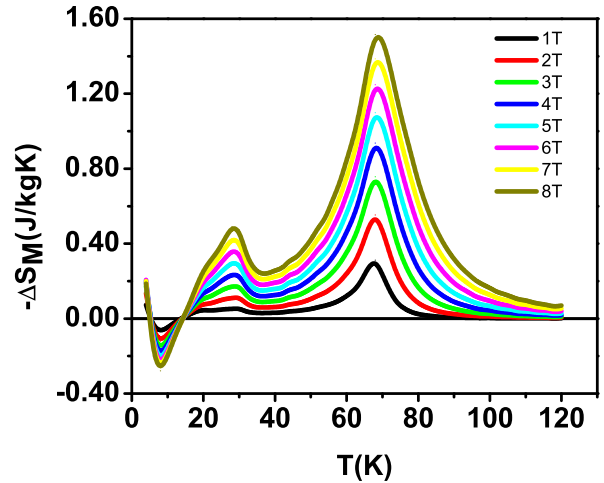}
\caption{Temperature dependence of MCE = $-\Delta S_M$ for various magnetic fields $H$.}
   	\label{MCE}
   \end{figure}

The closely spaced (every $2$~K) isothermal magnetization $M(H,T)$ data measured for NiCr$_2$O$_4$ are shown in Fig.~\ref{Series_M(H)} and the temperature dependent $-\Delta S_M(T, H)$ derived from these data is shown in Fig.~\ref{MCE}.  The $-\Delta S_M(T, H)$ exhibits positive anomalies at $T_c = 70$~K and $T_s = 30$~K corresponding to the anomalies observed in the magnetic measurements above. Another positive anomaly seen as a weak shoulder at $T_o \approx 20$~K is also observed.  This temperature is consistent with the recently reported anomaly in magnetocapacitance measurements \cite{19} and confirms that some magneto-structural change occurs at $T_o = 20$~K\@.  Additionally a negative anomaly (inverse MCE) is observed at $T \approx 8.5$~K\@.  This lower temperature anomaly hasn't been reported before and suggests additional magnetic or structural ordering or re-ordering of some kind.  

The transition at $T_c = 70$~K is reported to be a magneto-structural one.  In general, magnetostructural transitions are of first order transition in nature.  For a first order transition the order parameter emerges in a discontinuous manner due to which a large change in entropy is expected. The anomaly in $-\Delta S_M(T)$ at $T_c$ is however, quite smooth with a moderate value of $1.5$(J/kg-K) at $H = 8$~T and is symmetric around the $T_c$. This suggests that the transition at $T_c$ may be of second order.  The positive anomalies in MCE at $T_c$ and $T_s$ indicates decrease of magnetic entropy on the application of magnetic field.  This is expected due to spin re-orientation in a field.  The weak shoulder at $T = 20$~K also shows a magnetic field dependence which would be consistent with a magnetic transition.   

\subsection{Landau free energy and Arrott plot}
\begin{figure}
	\includegraphics[width=1.05\linewidth]{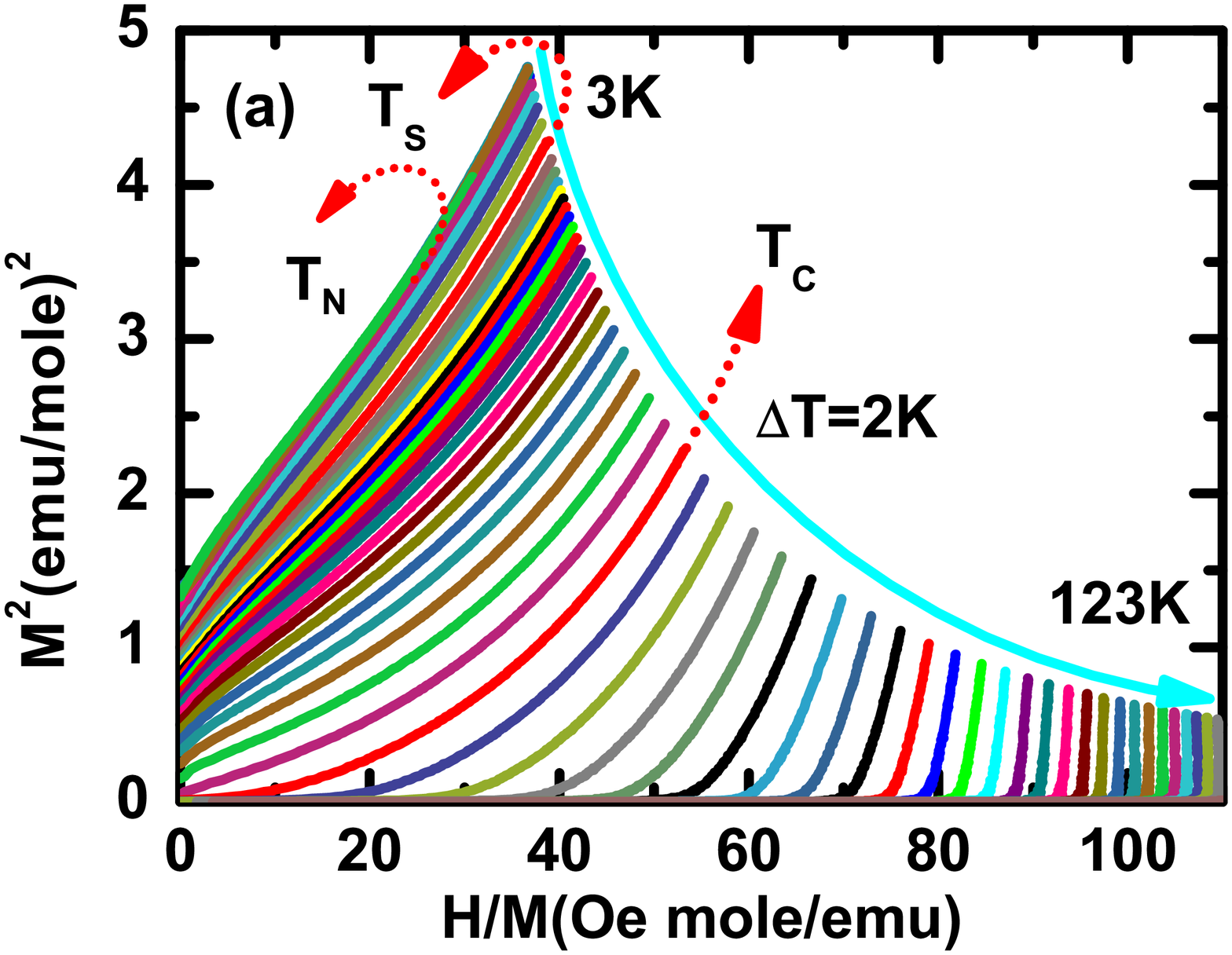}
	\includegraphics[width=1.05\linewidth]{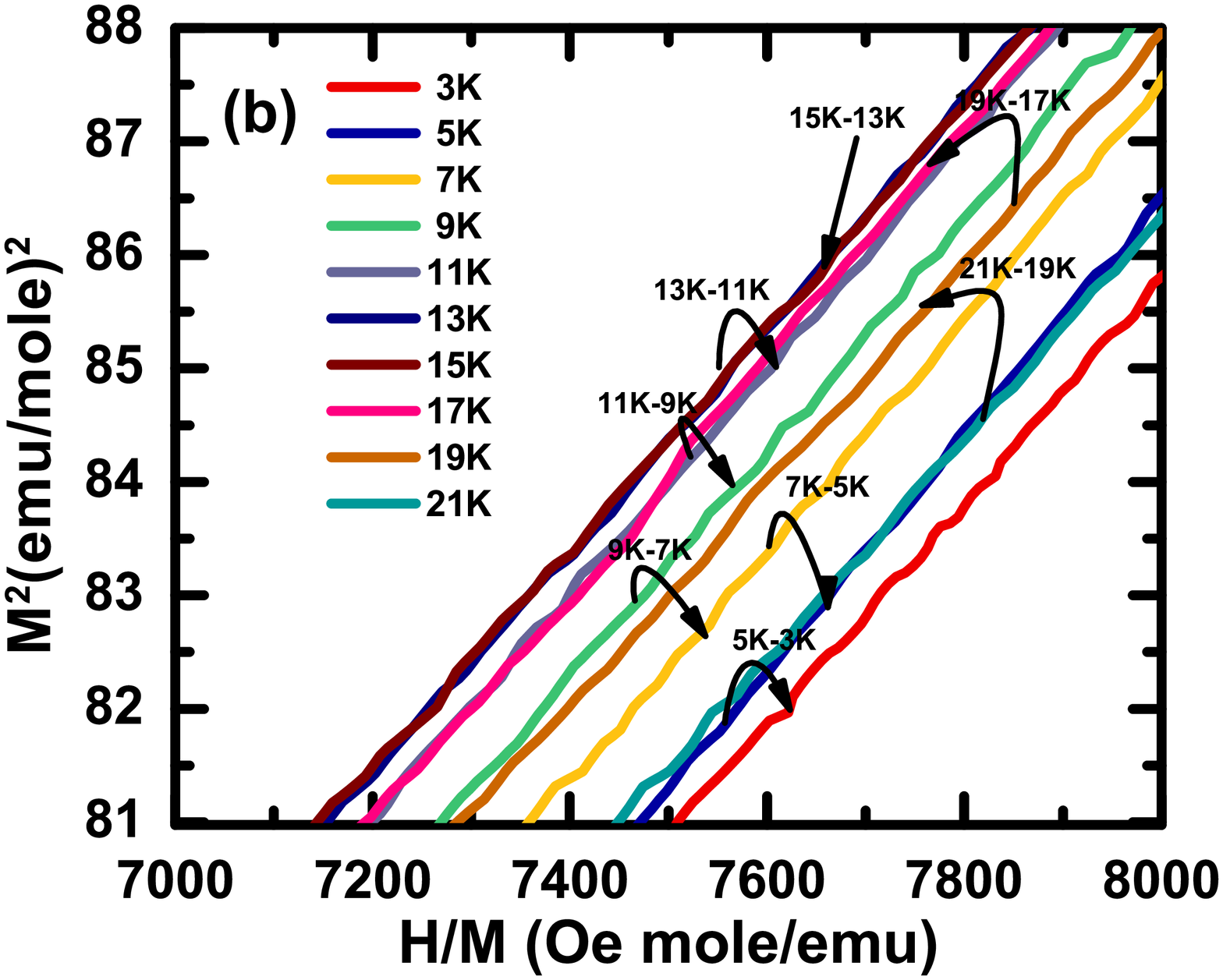}
	\caption{(a) Temperature dependence of Arrott plot derived from the isothermal curves of magnetization. The Arrott curve for the critical isotherm $T=T_c$ passes through the origin.  (b) The Arrott plots for $T \leq 21$~K on an expanded scale to highlight the progression of curves for various temperatures including reversal below $T \sim 11$~K  (see text for details). }
	\label{fig:ArrottPlot}
\end{figure} 

For magnetic phase transitions, the order parameter is the magnetization or sub-lattice magnetization if there are more than one sub-lattices.  The Landau free energy for a material like NiCr$_2$O$_4$ with different sub-lattice magnetizations $m_1$ and $m_2$ which are coupled, can be written as: 
\begin{equation}
F(m_1, m_2, H) = F_0 + \frac{1}{2} am_1^2 + \frac{1}{4} bm_1^4 -Hm_1 \nonumber
\end{equation}
\begin{equation}
+ \frac{1}{2}Am_2^2 + \frac{1}{4}Bm_2^4 -Hm_2 -Cm_1m_2~.
\label{free energy expression}
\end{equation}

This expression is constructed by assuming ferromagnetic intra-sublattice interaction and antiferromagnetic inter-sublattice interaction. The antiferromagnetic exchange between sublattices is taken into account by introducing the term $-Cm_1m_2$ in equation \ref{free energy expression} with $C < 0$.  As usual, the parameters $F_0$, $b > 0$ and $B > 0$ are taken as temperature independent constants while $a$ and $A$ are temperature dependent coefficients given by
               
\begin{eqnarray}
a(T) = a_0 + a_1T^2 > 0, \label{Thermal paramater_1}\\
A(T) = A_0 + A_1T^2 > 0. \label{Thermal Parameter_2}
\end{eqnarray}
  
To get solutions for ground state values of $m_1$ and $m_2$ we need to do a minimization of the free energy as $m_1$ and $m_2$ vary i.e. we need to set $\frac{\partial (F)}{\partial (m_1,m_2)} = 0$, which gives us:

\begin{eqnarray}
am_1 + bm_1^3 - H - Cm_2 = 0, \label{m1}\\
Am_2 + Bm_2^3 - H - Cm_1 = 0. \label{m2}
\end{eqnarray}

We see that the coupling term in the free energy leads to terms in Eqns.~\ref{m1} and \ref{m2} that act as effective magnetic fields.  For example, the magnetization of the second sublattice acts as an effective field for the first sublattice as can be seen in Eqn.~\ref{m1}.  The above expressions also indicate that $m_1$ and $m_2$ will be non-linear functions of $H$.  This will in turn lead to a non-linear Arrott plot ($M^2$ vs $H/M$) for the total magnetization $M$. 

The Arrott plot for NiCr$_2$O$_4$ obtained from the isothermal magnetization data in Fig. \ref{Series_M(H)} is shown in Fig. \ref{fig:ArrottPlot}~(a). The Arrott plot curves for different temperatures are non-linear and their curvature changes as the system is cooled across the various phase transitions. In the paramagnetic state, the curves are strongly concave, very different from the curves obtained for simple feromagnets which are linear.  Strong curvature in the Arrott curves for $T > T_c$ suggests strong antiferomagnetic coupling between the two (Cr$^{3+}$ and Ni$^{2+}$) sublattices.  The temperature for which the nonlinear Arrott curve passes the origin, is identified with the ferrimagnetic transition temperature $T_c$.  It can be seen from Fig.~\ref{fig:ArrottPlot}~(a) that this happens for the curve at $T = 70$~K which is consistent with anomalies in other measurements reported above. 

As the temperature is lowered below $T_c$ the concaveness of the Arrott curves reduces and eventually disappears and the curves show almost linear dependence for large $H/M$.  As we cross $T_s = 30$~K the Arrott curves do not show any significant change. This again suggests that the transition at $T_s$ is not magnetic in origin.  As the temperature is lowered further, something unusual happens.  The Arrott plots for temperature $T \leq 21$~K are shown in Fig.~\ref{fig:ArrottPlot}~(b) on an expanded scale.  We see that the Arrott curves, for temperatures down to $T = 17$~K are all moving up roughly by uniform amounts.  This can be seen from the Fig.~\ref{fig:ArrottPlot}~(b) for the plots going from $T = 21$~K to $T = 19$~K and going from $T = 19$~K to $T = 17$~K, which are shown as arrows.  Below $T = 17$~K, the rate at which the curves for different temperatures are moving up slows down and we se that the curve for $T = 15$~K is quite close to the $T = 17$~K curve.  The $T =13$~K curve almost overlaps with the $T = 15$~K curve and for temperatures lower than this, the curves reverse their progression and start moving down as $T$ is lowered below $T = 13$~K\@.  The rate at which the curves move down is increased at $T \approx 10$~K as can be seen from the change in the plots in going from $11$~K to $9$~K and $9$~K to $7$~K and so on.  This reversal in Arrott curves indicates that a new magnetic phase possibly of antiferromagnetic nature exists below $T \approx 10$~K \cite{33}.  This temperature is close to $T \approx 8.5$~K where the inverse MCE was observed.    

\subsection{Scaling analysis}

\begin{figure}
	\includegraphics[width=0.9\linewidth]{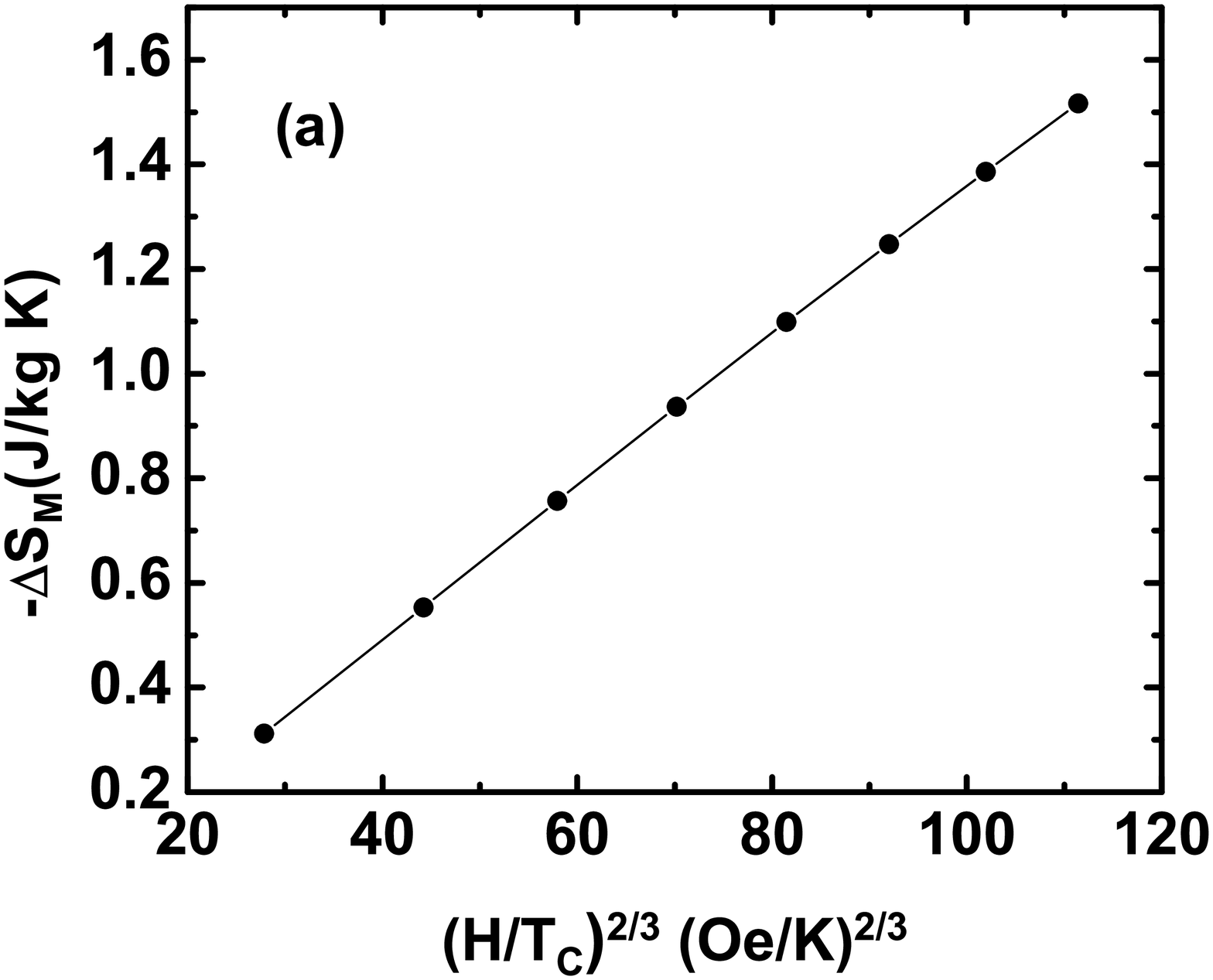}
	\includegraphics[width=0.9\linewidth]{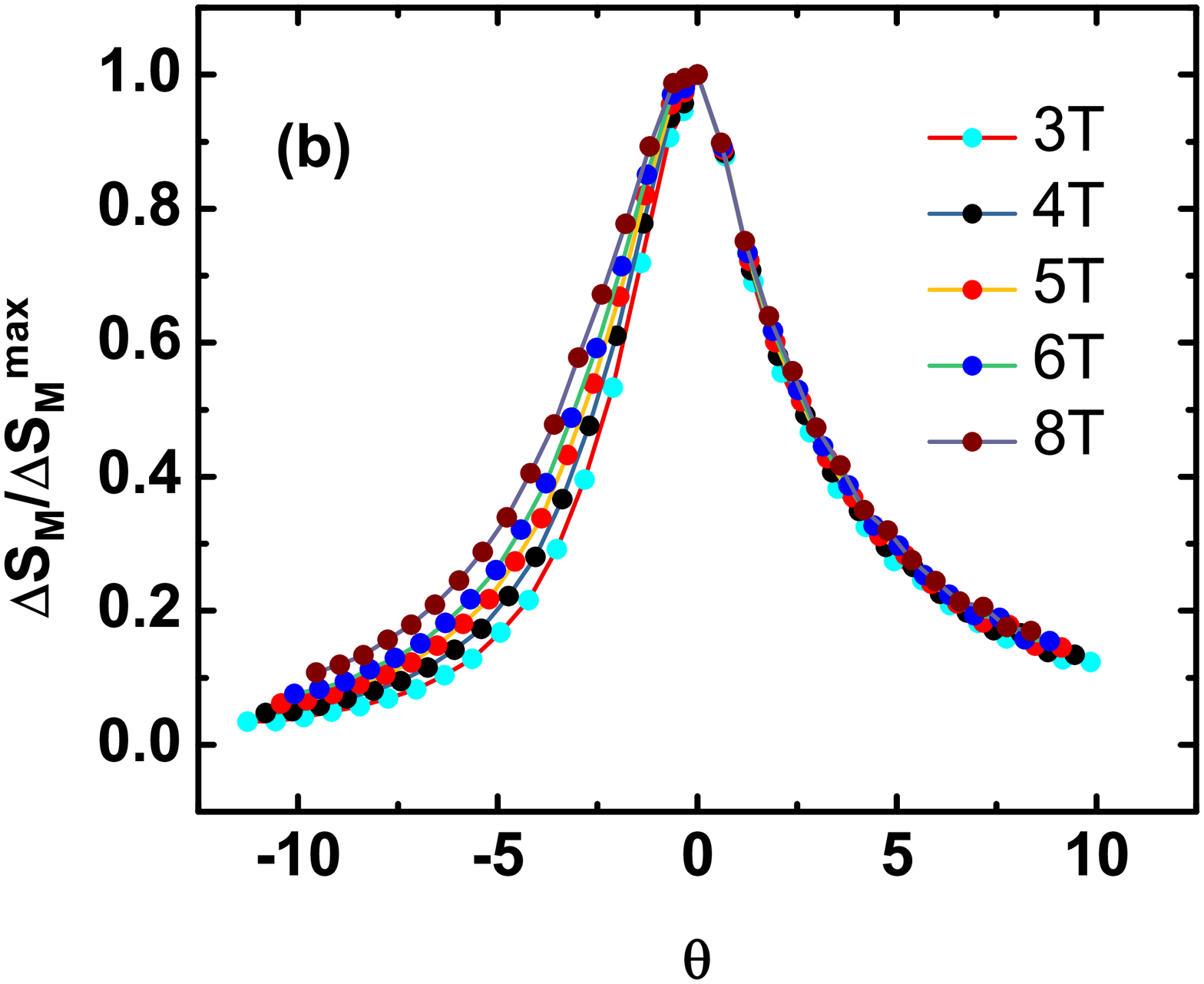}
	\caption{Scaling plots of $\Delta S_M$ for the $T = 70$~K transition.  (a) $\Delta S_M^{max}$ versus $H^{2/3}$ data showing a linear dependence expected for a mean-field transition.  (b) The collapse of all the $\Delta S_M$ data at different magnetic fields onto a universal curve when plotted as $\Delta S_M/\Delta S_M^{max}$ vs $\theta$ (see text for details).}
	\label{scaling-70K}
\end{figure}

\begin{figure}
	\includegraphics[width=0.9\linewidth]{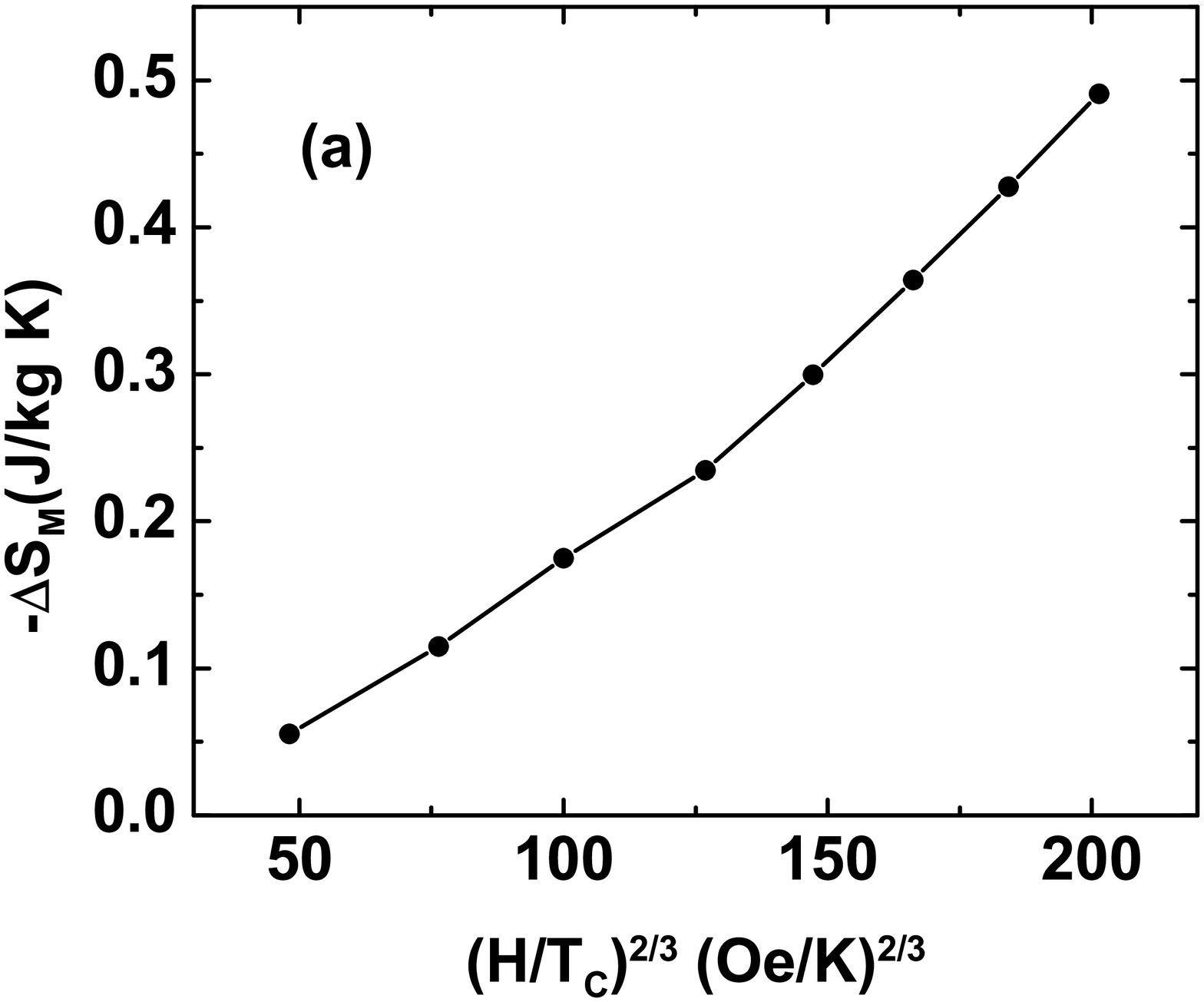}
	\includegraphics[width=0.9\linewidth]{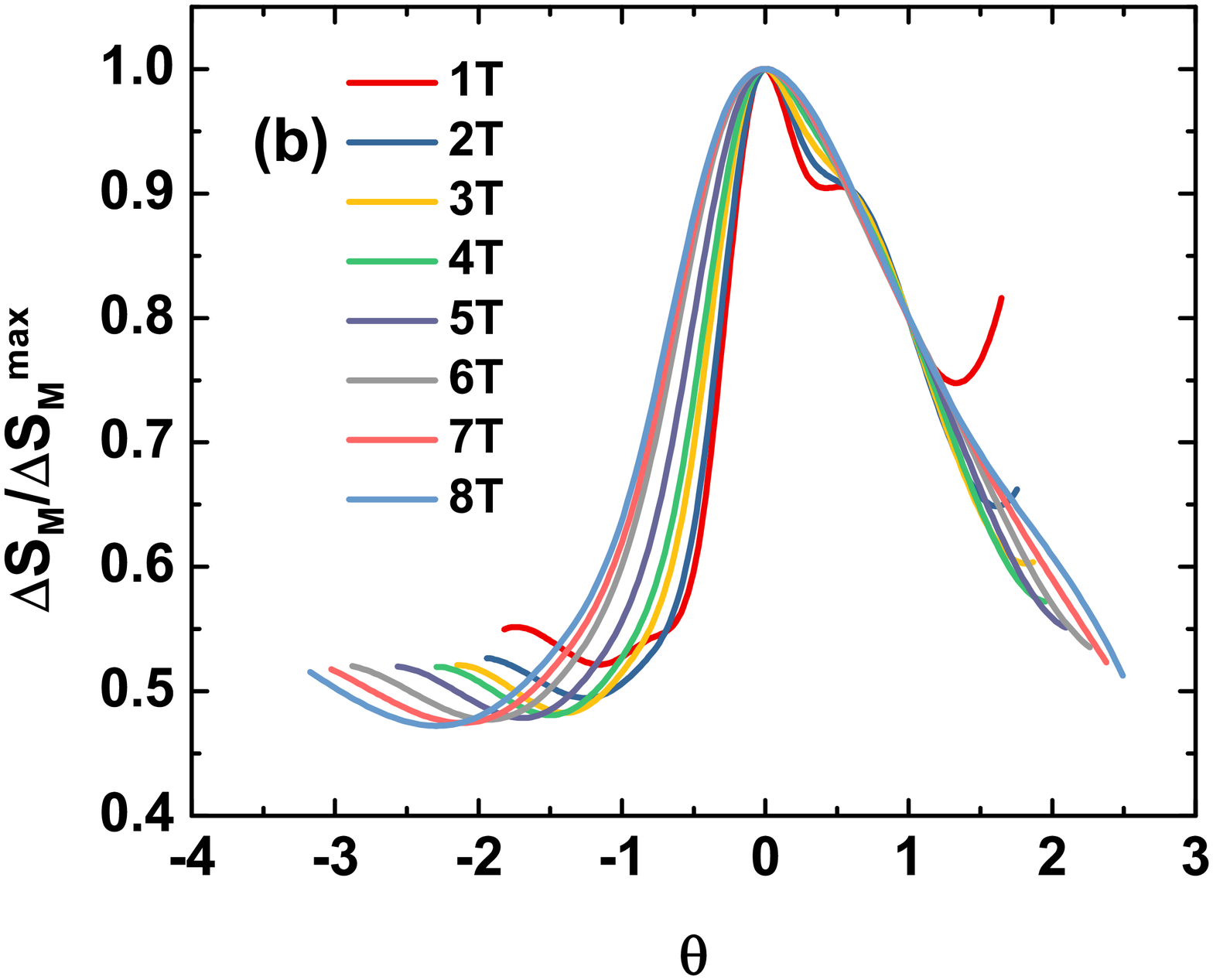}
	\caption{Scaling plots of $\Delta S_M$ for the $T = 30$~K transition.  (a) $\Delta S_M^{max}$ versus $H^{2/3}$ data showing a non-linear dependence.  (b) The $\Delta S_M/\Delta S_M^{max}$ vs $\theta$ at various magnetic fields do not collapse to a universal curve (see text for details). }
	\label{scaling-30K}
\end{figure}

\begin{figure}
	\includegraphics[width=0.9\linewidth]{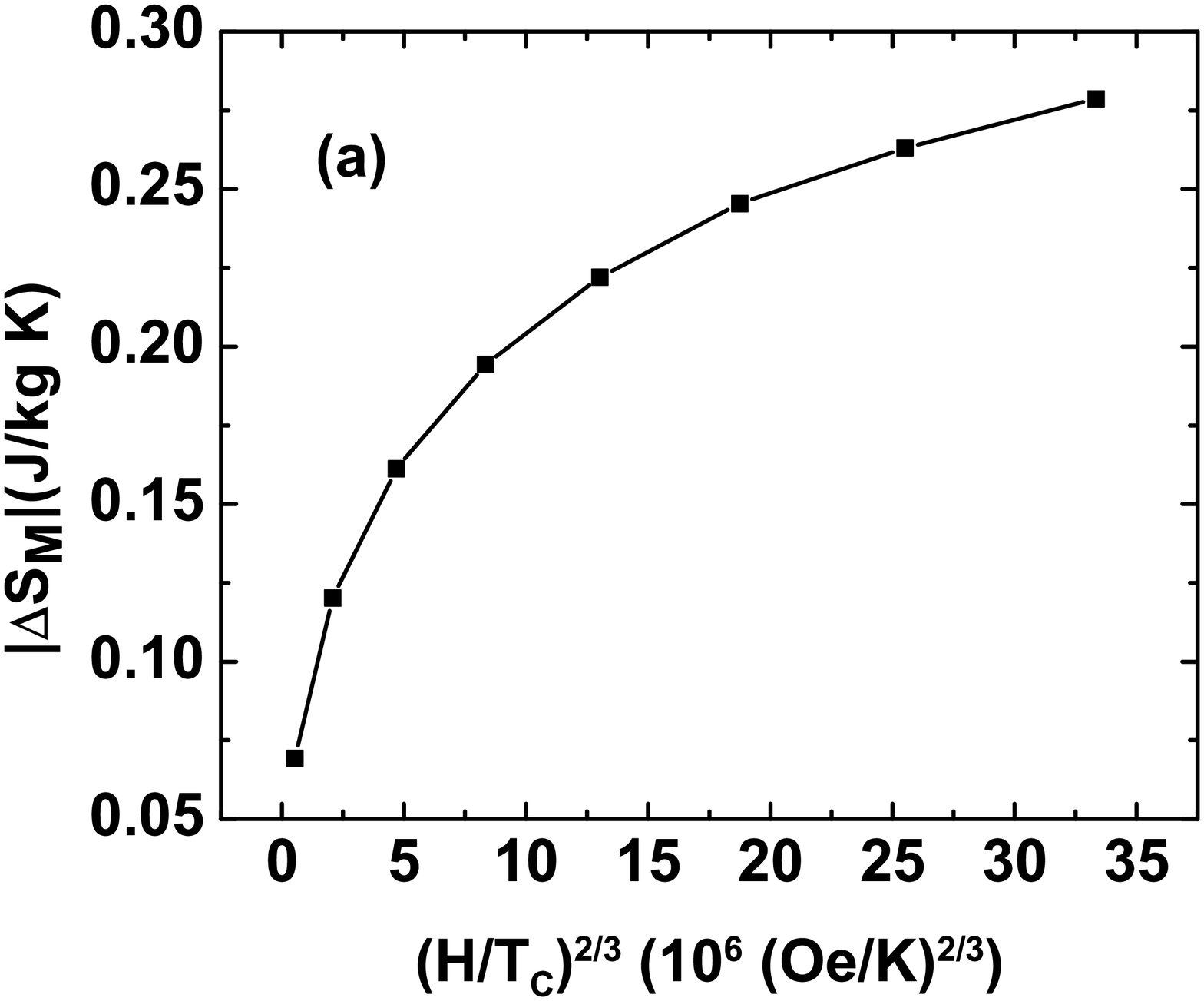}
	\includegraphics[width=0.9\linewidth]{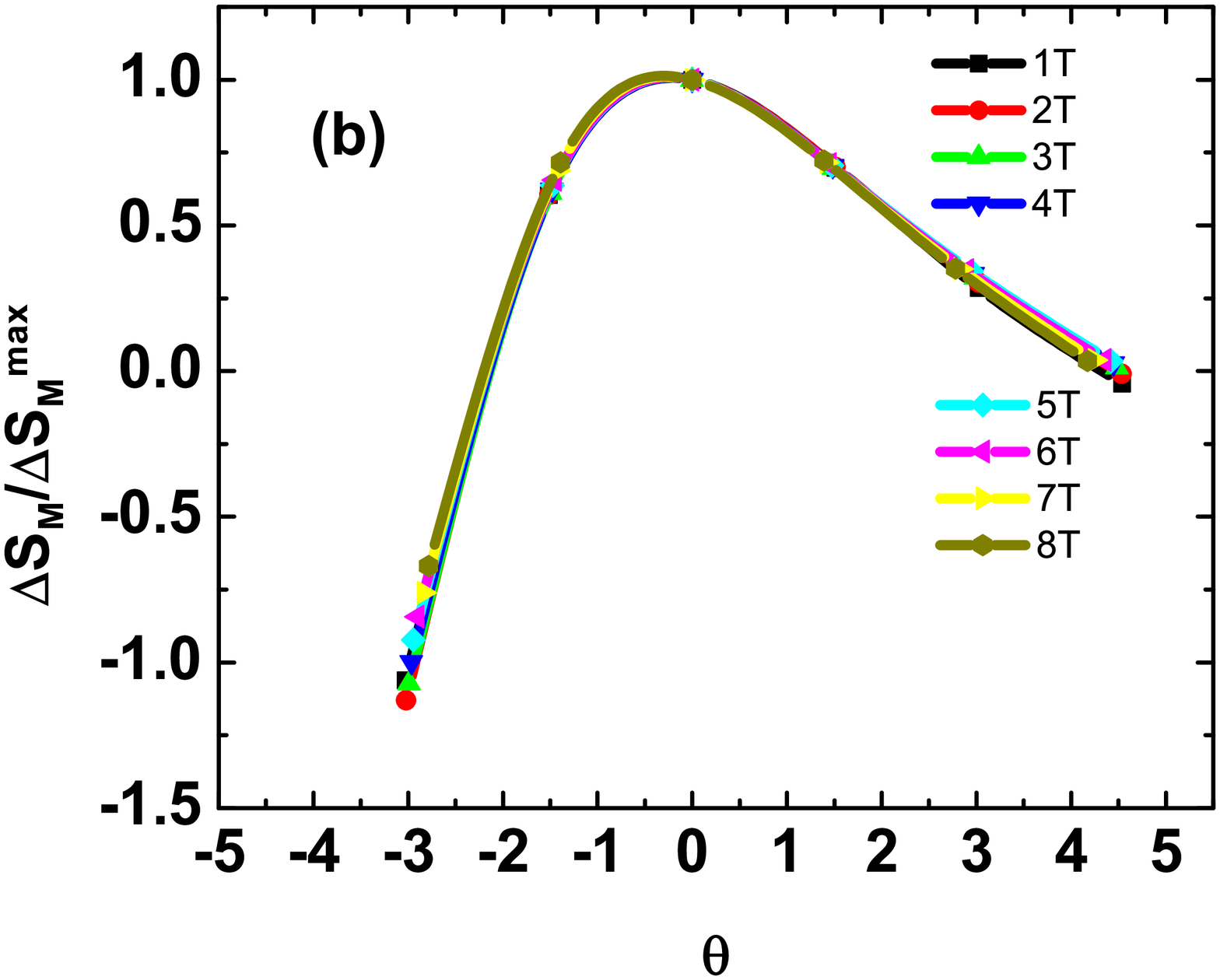}
	\caption{Scaling plots of $\Delta S_M$ for the $T = 8.5$~K inverse MCE peak.  (a) $\Delta S_M^{max}$ versus $H^{2/3}$ data showing a strongly non-linear dependence.  (b) The $\Delta S_M/\Delta S_M^{max}$ vs $\theta$ at various magnetic fields collapse to a universal non-Lorentzian curve (see text for details).}
	\label{scaling-8K}
\end{figure}

Within mean field theory the magnetic entropy change at the transition is expected to follow a power law behaviour given by $\Delta S_M \propto H^n$ with $n = 2/3$~ \cite{2}.  Additionally, close to a second order phase transition the $\Delta S_M(T)$ curves at different magnetic fields are expected to collapse onto a common universal curve when they are plotted as $\Delta S_M/\Delta S_M^{max}$ versus $\theta$, where $\Delta S_M^{max}$ is the value of $\Delta S_M$ at the transition temperature around which the scaling analysis is being made, and $\theta$ is a reduced temperature given by $\theta = -{T - T_c \over T_{r1} - T_c}$ for $T \leq T_c$ and $\theta = {T - T_c \over T_{r2} - T_c}$ for $T > T_c$.  The $T_{r1}$ and $T_{r2}$ are the temperatures at the full width at half maximum of the anomaly in $\Delta S_M$~\cite{2}.  We have performed the above scaling analysis for the anomalies observed at $T_c = 70$~K, $T_s = 30$~K, and at $T = 8.5$~K in the $\Delta S_M(T)$ data shown in Fig.~\ref{MCE}.  

For the ferrimagnetic transition at $T_c = 70$~K, the plot of $\Delta S_M$ versus $H^{2/3}$ is shown in Fig.~\ref{scaling-70K}~(a).  A linear plot is obtained which strongly suggests that this transition is mean-field like.   A plot of $\Delta S_M/\Delta S_M^{max}$ at various fields versus the reduced parameter $\theta$ is shown in Fig.~\ref{scaling-70K}~(b).  We see that all the $\Delta S_M$ curves for the different magnetic fields approximately collapse onto a single universal master curve.  This is strong evidence of the second order nature of the ferrimagnetic transition at $T_c = 70$~K in NiCr$_2$O$_4$.   

For the structural transition at $T_s = 30$~K, the $\Delta S_M$ versus $H^{2/3}$ shown in Fig.~\ref{scaling-30K}~(a) is non-linear indicating a non-mean-field like transition.  A plot of $\Delta S_M/\Delta S_M^{max}$ versus $\theta$ shown in Fig.~\ref{scaling-30K}~(b) is less conclusive.  It seems that the $\Delta S_M$ curves at different fields do collapse somewhat to a single curve for $\theta > 0$.  However, for $\theta < 0$ there is clear separation of the curves at different magnetic fields suggesting no collapse. We therefore conclude from this overall non-universal scaling to mean that the transition at $T_s = 30$~K is not a second order transition.

Finally, for the $T = 8.5$~K inverse MCE anomaly, the $\Delta S_M$ versus $H^{2/3}$ shown in Fig.~\ref{scaling-8K}~(a) is clearly non-linear indicating a non-mean-field like transition.  A plot of $\Delta S_M/\Delta S_M^{max}$ versus $\theta$ is shown in Fig.~\ref{scaling-8K}~(b).  We see that all the $\Delta S_M$ curves for the different magnetic fields beautifully collapse onto a single universal curve.  This is strong evidence that the new transition at $T = 8.5$~K suggested from the strong inverse MCE, might be second order in nature.  We note that while the universal MCE curve usually has a Lorentzian shape \cite{2}, also seen for the $T = 70$ and $30$~K transitions, the universal $\Delta S_M$ curve for the $8.5$~K transition is non-Lorentzian.    

\section{Conclusion}
The spinel NiCr$_2$O$_4$ is a magneto-elastic material where the lattice degree of freedom is strongly coupled with spin and orbital degrees of freedom. This coupling drives several temperature dependent phase transitions which are magnetic or magneto-structural in nature.  

In this work we study the magnetocaloric (MCE) response $-\Delta S _M$ of NiCr$_2$O$_4$ as the temperature is varied across the various magnetic, structural or coupled magneto-structural transitions.  The MCE shows three positive anomalies at $T_c = 70$~K, $T_s = 30$~K, and $T_o \approx 20$~K\@.  The first two anomalies have been well characterized in literature while the third one has been reported only in one previous work.  Our work therefore confirms that some phase change occurs below $T_o = 20$~K\@.  Additionally, the MCE shows a negative anomaly at $\approx 8.5$~K\@.  This robust inverse MCE anomaly points to a hitherto unreported transition which is most likely magnetic in nature.  Future microscopic probe studies will be required to clarify the origin of this possible new phase transition in NiCr$_2$O$_4$.  A scaling analysis of the $\Delta S_M$ data near the various critical temperatures reveals that the transition at $T_c$ is a second-order Mean field like transition, the transition at $T_s$ is not second order and is non-mean field like, while the newly discovered transition at $T = 8.5$~K is non-mean field like but is second order in nature.

\section{Acknowledgment}
We thank the X-ray facility at IISER Mohali.  AA thanks MHRD for financial support.
 
\bibliographystyle{apsrev4-1}
\bibliography{Ref.bib}
\end{document}